\documentclass[11pt,a4paper]{article}
\usepackage{jcappub}

\usepackage{amsmath,amsthm,latexsym,amssymb,amsfonts,epsfig}
\usepackage{fontenc,dsfont,layout}
\usepackage{hyperref}
\usepackage{psfrag}
\usepackage[cp1250]{inputenc}
\usepackage{graphicx}

\usepackage[usenames,dvipsnames]{xcolor}

\numberwithin{equation}{section}

\title{Inflation and dark energy from \textit{f}(\textit{R}) gravity}
\author[a]{Micha{\l}~Artymowski}
\author[b]{Zygmunt~Lalak}
\affiliation[a]{Department of Physics, Beijing Normal University\\
Beijing 100875, China}
\affiliation[b]{Institute of Theoretical Physics, Faculty of Physics, University of Warsaw\\ 
ul. Ho\.{z}a 69, 00-681 Warszawa, Poland}

\emailAdd{artymowski@bnu.edu.cn}
\emailAdd{Zygmunt.Lalak@fuw.edu.pl}

\abstract{The standard Starobinsky inflation has been extended to the  $R + \alpha R^n -
\beta R^{2-n}$ model to obtain a stable minimum of the Einstein frame scalar
potential of the auxiliary field. As a result we have obtained obtain a scalar
potential with non-zero value of residual vacuum energy, which may be a source of
Dark Energy. Our results can be easily consistent with PLANCK or BICEP2 data for appropriate choices of the value of $n$.}


\arxivnumber{1405.7818}

\begin{document}
\maketitle

\section{Introduction}

Inflationary scenario for the Early Universe has become one of the standard assumptions 
in theoretical model building. It can account naturally for the wealth of present observational data. The existence of rapid expansion stage in the history of the Universe can  give us a natural
explanation of the homogeneity, flatness and horizon  problems, and provides an explanation  of  the process of seeding  large-scale structure and temperature anisotropies of the cosmic microwave background radiation (see refs. \cite{LR99,Liddle,MR11} for reviews).
\\*

One of the first inflationary models was the Starobinsky inflation \cite{Starobinsky:1980te}, based on the $f(R) = R + R^2/6M^2$ Lagrangian density.  In this theory during inflation the $R^2$ term dominates, which provides a stage of the de-Sitter-like evolution of space-time. The inflationary potential has a stable minimum, which allows for the graceful exit, reheating and good low-energy limit of the theory. However, the recent observational data from the BICEP2 experiment \cite{Ade:2014xna,Ade:2014gua} suggest significant amount of primordial gravitational waves, which disfavours the Starobinsky inflation. As shown in \cite{Codello:2014sua}, a small modification of the $f(R)$ function, namely $f(R) = R + \alpha R^n$, can generate inflation which fits the BICEP2 data. This model has already been partially discussed in \cite{DeFelice:2010aj}. Different modifications of the Starobinsky model were also discussed in Ref. \cite{Ben-Dayan:2014isa,Sebastiani:2013eqa}.
\\*

On the other hand the series of experiments \cite{Perlmutter:1999jt,Spergel:2006hy,Ade:2013zuv} convincingly suggests the existence of the so-called Dark Energy (DE) with barotropic parameter close to $-1$. One of the possible sources of DE may be  a non-zero vacuum energy of a scalar field, which in principle can be the $f(R)$ theory \cite{DeFelice:2010aj,Capozziello:2002rd,Copeland:2006wr} or the Brans-Dicke field. 
\\*

In this paper we demonstrate how to extend the $f(R) = R + \alpha R^n$ model to combine successfully both features of the observable Universe in a single framework: we show how to obtain successful inflation and the non-zero residual value of the Ricci scalar in an extension of the Starobinsky model.  Namely, we use the $f(R)=\alpha R^n - \beta R^{2-n}$ Lagrangian density, which provides inflation from the $\alpha R^n$ and a stable minimum of the scalar potential of an auxiliary field. The potential has a non-zero vacuum energy density, which is the source of the DE in the late-times evolution. Let us note that some inflationary potentials, which generate DE (motivated by modified theories of gravity) were already introduced in the Ref. \cite{Nojiri:2003ft}
\\*

In the following draft we use the convention $8\pi G = M_{pl}^{-2} = 1$, where $M_{pl}\sim 2\times 10^{18}GeV$ is the reduced Planck mass.
\\*

The structure of this paper is as follows. In the Sec. \ref{sec:R^n} we analyse the $f(R) = R + \alpha R^n$ model: we discuss analytic  solutions, parameters of inflation and primordial inhomogeneities. In the Sec. \ref{sec:R^1+m} we generalise this model to $f(R) = R + \alpha R^n - \beta R^{2-n}$, which leads to the non-zero vacuum energy of the Einstein frame potential. In the Sec. \ref{sec:DE} we discuss numerical study of the evolution of the model with dust modelling the contribution of  matter fields to energy density. Finally, we conclude in the Sec. \ref{sec:concl}.

\section{The $R+\alpha R^n$ model} \label{sec:R^n}

Let us consider an $f(R)$ theory in the flat FRW space-time with the metric tensor of the form of $ds^2 = -dt^2 + a(t)^2(d\vec{x})^2$. Then, the Friedmann equations become
\begin{eqnarray}
3FH^{2} &=& (FR-f)/2-3H\dot{F}+ \rho_M\,,\\ 
-2F\dot{H} &=& \ddot{F}-H\dot{F}
+ (\rho_M+P_M)\,, \label{eq:Fried2}
\end{eqnarray}
where
$$ F(R) = \frac{df}{dR} $$
and $\rho_M$ and $P_M$ are energy density and pressure of matter fields respectively. 
\\*

The action of $f(R)$ model can be rewritten as
\begin{equation}
S = \int d^4x \sqrt{-g} \left[ \frac{1}{2}\varphi R - U(\varphi) + \mathcal{L}_{\text{m}}\right]\ ,
\end{equation}
where 
\begin{equation}
\varphi = F(R)\ ,\qquad U(\varphi) = \frac{1}{2}(RF-f) \, ,
\end{equation}
which means that the $f(R)$ gravity can be expressed as a Brans-Dicke theory in Jordan frame with $\omega_{BD} = 0$. In the Jordan frame the auxiliary field $\varphi$ is non-minimally coupled to gravity, which creates a deviation from the General Relativity (GR) frame. The first Friedmann equation and continuity equations are the following ones
\begin{eqnarray}
\ddot{\varphi} + 3H\dot{\varphi} + \frac{2}{3}(\varphi U_\varphi-2U) &=& \frac{1}{3}(\rho_M-3p_M)\, ,\label{eq:varphiEOM}\\
3\left(H+\frac{\dot{\varphi}}{2\varphi}\right)^2 &=& \frac{3}{4}\frac{\dot{\varphi}^2}{\varphi^2} + \frac{U}{\varphi} + \frac{\rho_M}{\varphi}\, ,\label{eq:FriedBD}\\
\dot{\rho}_M + 3H(\rho_M + P_M) &=& 0\, . \label{eq:cont}
\end{eqnarray}
where $U_\varphi=\frac{dU}{d\varphi}$. Let us note that $U$ may be interpreted as a energy density, but $U_\varphi$ is not an effective force in the Eq. (\ref{eq:varphiEOM}). One can define the effective potential and its derivative - the effective force, by
\begin{equation}
U_{eff} = \int \frac{2}{3}(\varphi U_\varphi-2U) d \varphi - \varphi  \frac{1}{3}(\rho_M-3p_M) + C\, , \qquad F_{eff} := \frac{d U_{eff}}{d\varphi}\, , 
\end{equation}
where $C$ is unknown constant of integration. The effective potential shall be interpreted as a source of an effective force, but not as a energy density.
\\*

The gravitational part of the action may obtain its canonical (minimally coupled to $\varphi$) form after transformation to Einstein frame
\begin{equation}
\tilde{g}_{\mu\nu}=\varphi g_{\mu\nu}\, , \qquad d\tilde{t}=\sqrt{\varphi}dt\, ,\qquad\tilde{a} = \sqrt{\varphi}a
\end{equation}
which gives the action of the form of 
\begin{equation}
S = \int d^4x \sqrt{-\tilde{g}}\left[ \frac{1}{2}\tilde{R} - \frac{3}{4}\left( \frac{\tilde{\nabla}\varphi}{\varphi} \right)^2 - \frac{U(\varphi)}{\varphi^2} \right]\, ,
\end{equation}
where $\tilde{\nabla}$ is the derivative with respect to $\tilde{x}^\mu$. In order to obtain the canonical kinetic term for $\varphi$ let us use the Einstein frame scalar field $\phi$ 
\begin{equation}
\phi = \sqrt{\frac{3}{2}}\log\varphi \, .
\end{equation}
The action in terms of $\tilde{g}_{\mu\nu}$ and $\phi$ looks as follows
\begin{equation}
S = \int d^4x \sqrt{-\tilde{g}}\left[ \frac{1}{2}\tilde{R} - \frac{1}{2}\left( \tilde{\nabla}\phi \right)^2 - V(\phi)\right]\, ,
\end{equation}
where
\begin{equation}
V = \frac{(FR-f)}{2F^2} = \left.\frac{U(\varphi)}{\varphi^2}\right|_{\varphi=\varphi(\phi)}\, .
\end{equation}
Let us note that $(\varphi U_\varphi-2U)$ from the Eq. (\ref{eq:varphiEOM}) can be expressed as $\varphi^3V_\varphi$, so the minimum of $V$ shall also be the minimum of the effective potential in the Jordan frame. In fact, all important features of the potential, like existence of minima and barriers between them, which determine the evolution of the field in the Einstein frame are reflected in  the evolution of the field in the Jordan frame. In further parts of this draft we will refer to the Einstein frame potential, even though we consider the Jordan frame as the primordial (or defining) one. Since all of the analysis performed in this draft is classical, descriptions in both frames give the same physical results. However, the description in the Einstein  frame is more intuitive, due to the canonical form of the scalar field's kinetic term and the minimal coupling between the field and the gravity. The only exception is the $\varphi\to\infty$ limit, which usually leads to $V\to\infty$ due to the $\varphi^{-2}$ term in the potential. This infinity comes from the singularity of the Einstein frame metric tensor and it does not appear in the Jordan frame analysis.
\\*

\begin{figure}[h]
\centering
\includegraphics[height=5.6cm,bb=0 0 288 222]{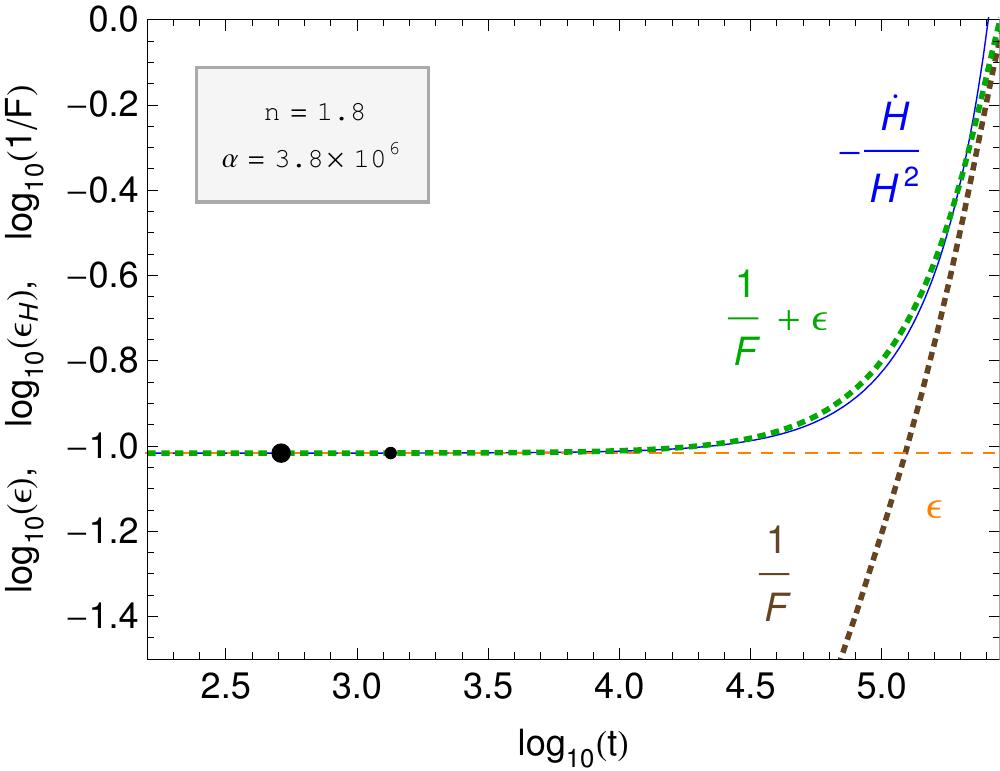}
\hspace{0.5cm}
\includegraphics[height=5.6cm,bb=0 0 288 222]{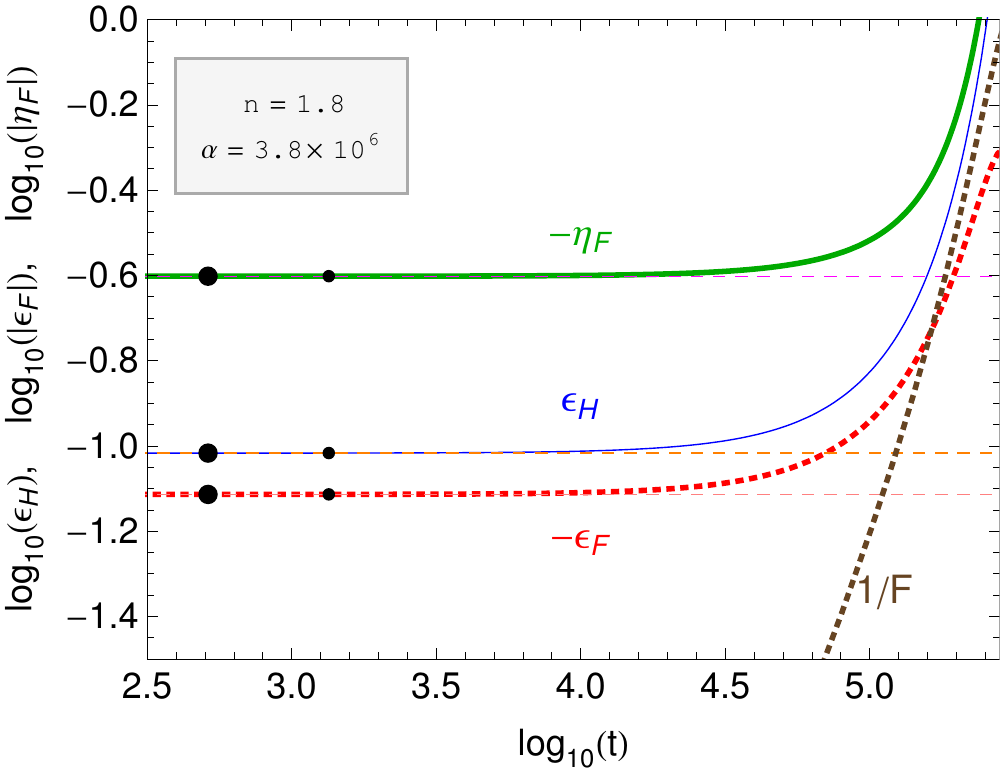}
\caption{\it Analytical results versus numerical simulation of slow-roll parameters and their $\epsilon F>1$ limit. Dots correspond to $N=60$ and $N=50$ (left and right dots respectively). At the moment of the horizon crossing one obtains $\epsilon F\gg1$, so one can use the analytical solution from the Eq. (\ref{eq:epsilonconstant}) to describe the evolution of space-time during that period.}
\label{fig:epsilon1_8}
\end{figure}

\begin{figure}[h]
\centering
\includegraphics[height=5.6cm,bb=0 0 288 222]{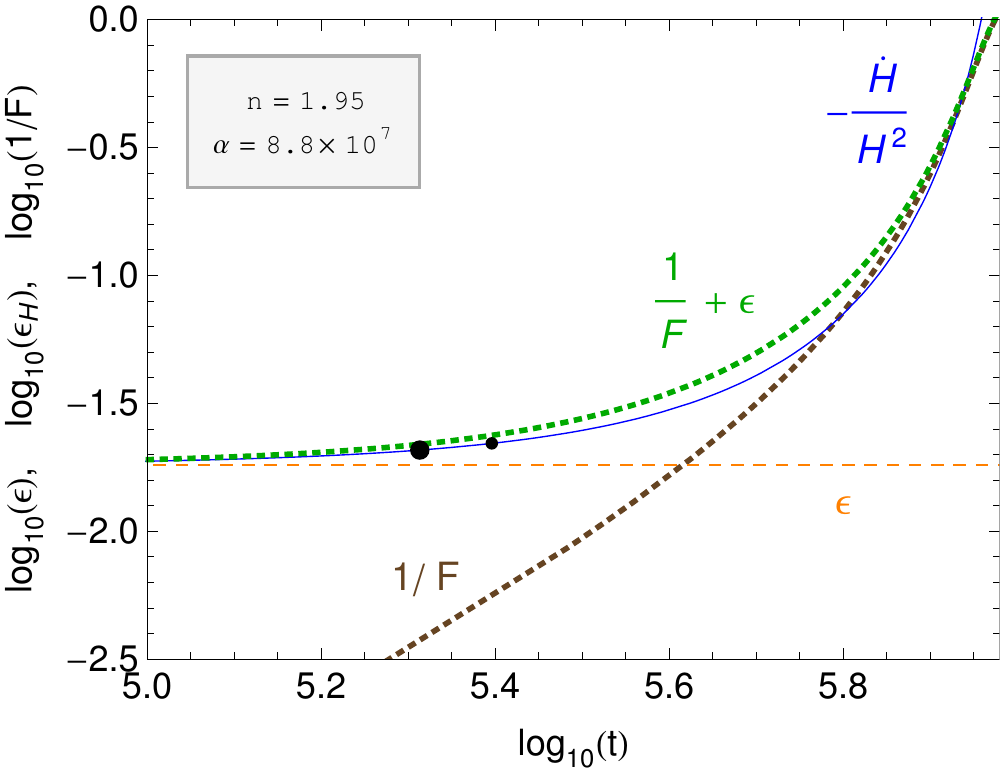}
\hspace{0.5cm}
\includegraphics[height=5.6cm,bb=0 0 288 222]{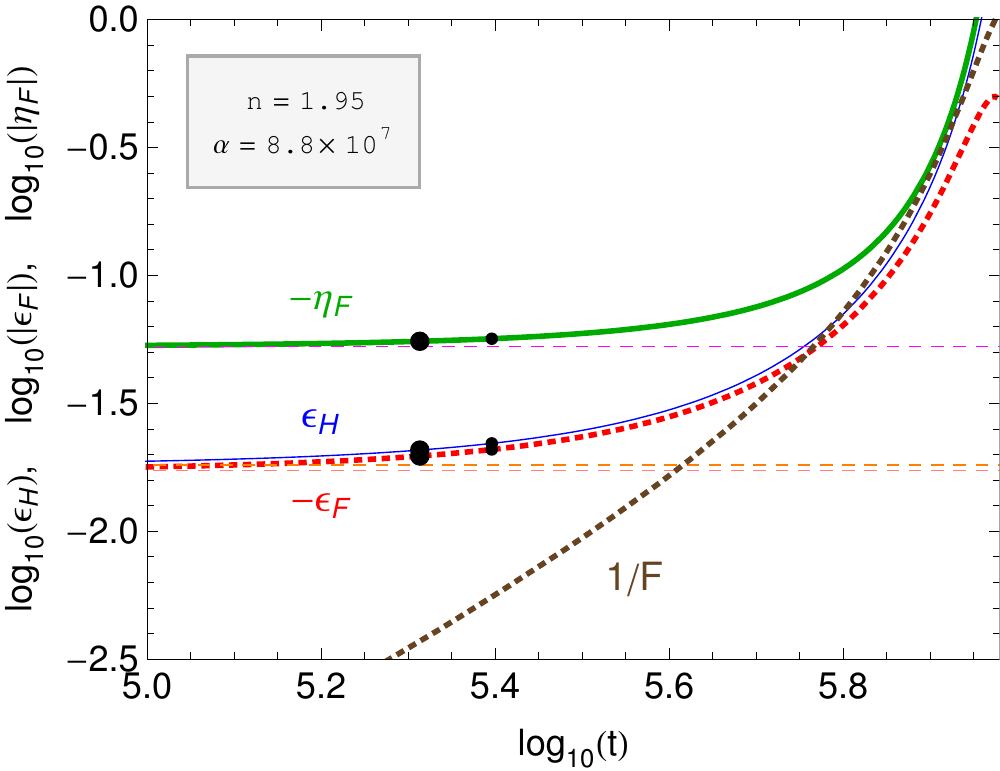}
\caption{\it Analytical results versus numerical simulation of the slow-roll parameters and their $\epsilon F>1$ limit. Dots correspond to $N=60$ and $N=50$ (left and right dots respectively). At the moment of the horizon crossing the condition $\epsilon F>1$ is satisfied and one can use the analytical solution from the Eq. (\ref{eq:epsilonconstant}) to describe the evolution of space-time during that period. }
\label{fig:epsilon1.95}
\end{figure}

\begin{figure}[h]
\centering
\includegraphics[height=5.6cm,bb=0 0 288 222]{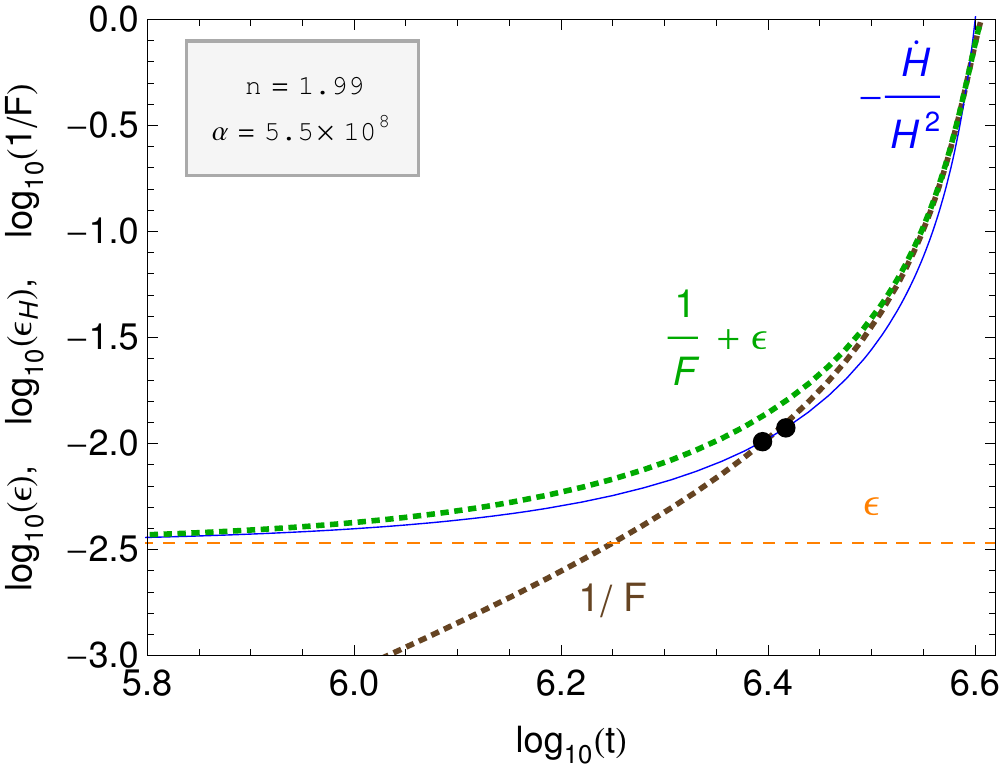}
\hspace{0.5cm}
\includegraphics[height=5.6cm,bb=0 0 288 222]{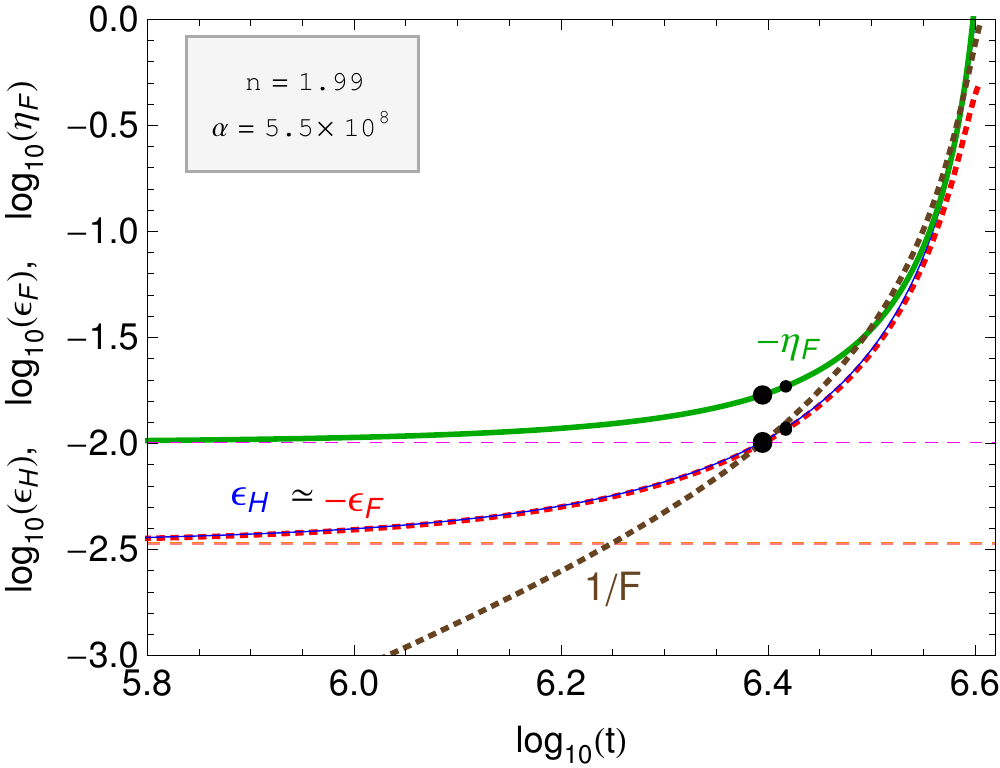}
\caption{\it Analytical results versus numerical simulation of the slow-roll parameters. Dots correspond to $N=60$ and $N=50$ (left and right dots respectively). At the moment of the horizon crossing one obtains $\epsilon F\ll1$, so the analytical solution from the Eq. (\ref{eq:epsilonconstant}) cannot be used to describe the evolution of space-time during that period.}
\label{fig:epsilon1.99}
\end{figure}

For the vacuum model with
\begin{equation}
f(R)=R+\alpha R^n\,,\qquad
(\alpha>0, n>0)\, ,\label{eq:fR}
\end{equation}
one obtains
\begin{equation}
3H^2(1+n \alpha R^{n-1})=\frac{1}{2} (n-1)\alpha R^n
-3n(n-1)\alpha H R^{n-2} \dot{R}\,.
\label{eq:FiredFRW}
\end{equation}
In the regime $F\gg1$, $\epsilon F>1$ one finds \cite{DeFelice:2010aj}
\begin{equation}
\frac{\dot{H}}{H^2} \simeq  -\epsilon\,, \qquad
\epsilon = \frac{2-n}{(n-1)(2n-1)}\,, \qquad H \simeq \frac{1}{\epsilon t}\,, \qquad
a \propto t^{1/\epsilon}\,. \label{eq:epsilonconstant}
\end{equation}
The same results would be obtained for the power-low inflation in GR frame (for minimally coupled scalar field) with the potential $V = V_0 e^{\lambda\phi}$, where $\epsilon = \lambda^2/2$. However, such a model would generate different spectrum of primordial inhomogeneities. Let us note that the $\epsilon$ can be interpreted as a slow-roll parameter only for $F\gg1$ and $\epsilon>1/F$. The second condition comes from the fact, that the correction to the equation of motion coming from the GR low-energy limit shall be of the order of $1/F\sim R/f(R)$. The slow-roll parameter will be denoted as $\epsilon_H$ end defined by
\begin{equation}
\epsilon_H:= -\frac{\dot{H}}{H^2} \simeq \epsilon + \frac{1}{F} \, .
\end{equation}
The inflation ends when $\epsilon_H\sim 1$, which happens in the $\epsilon F<1$ regime. Thus at the moment of the end of inflation one finds $\varphi\simeq 1$ for all realistic values of $n$.  Besides $\epsilon_H$ let us define slow-roll parameters $\epsilon_F$ and $\eta_F$ by
\begin{equation}
\epsilon_F := \frac{\dot{F}}{2HF}\, ,\qquad \eta_F := \frac{\ddot{F}}{H\dot{F} }\, . \label{eq:SR}
\end{equation}
To obtain inflation one must satisfy $\epsilon_H, \epsilon_F, \eta_F\ll1$. To satisfy this condition in the $\epsilon F>1$ limit we need
$$ \epsilon<1\quad\Leftrightarrow \quad n > \frac{1}{2}(1+\sqrt{3}) \, .$$
To satisfy $\dot{H}<0$ and $H>0$ one needs $n<2$. Thus the allowed range for $n$ is given by \cite{DeFelice:2010aj}
\begin{equation}
n \in \left[\frac{1}{2}(1+\sqrt{3}),2\right]\, . \label{eq:nin}
\end{equation}
Let us check whether the assumption $F\gg1$ is valid during the last 50-60 e-folding of inflation, during which the primordial inhomogeneities were generated. The evolution of a slow-roll parameters (and their $\epsilon F>1$ limit) as a function of time for different values of $n$ has been presented at fig. \ref{fig:epsilon1_8}, \ref{fig:epsilon1.95}, \ref{fig:epsilon1.99}. Values of the $\alpha$ parameter have been chosen to satisfy normalisation of primordial curvature perturbations at $50-60$ e-folds before the end of inflation.
\\*

To obtain the number of e-folds let us note that for the slow-roll approximation {of Eq. (\ref{eq:varphiEOM},\ref{eq:FriedBD},\ref{eq:fR})} one finds 
\begin{equation}
N\simeq\tilde{N}\simeq -\frac{3}{4} \log\left(\varphi\right)+\frac{3 n}{4(2-n)} \log\left(\frac{\varphi  (2-n)+2 (n-1)}{n}\right)\, ,\label{eq:NEinJord}
\end{equation}
where $N$ and $\tilde{N}$ are the number of e-folds in Jordan and  Einstein frames respectively. Usually the $\log(\varphi)$ term is subdominant and one can write
\begin{equation}
\varphi(N) \simeq \frac{1}{2-n}\left(2-\left(2-e^{\frac{4 (2-n)}{3 n}N}\right) n\right) \, .
\end{equation}
This result has been obtained using the slow-roll approximation and its accuracy is of order of few e-foldings.
\\*

\begin{figure}[h]
\centering
\includegraphics[height=3.7cm,bb=0 0 288 218]{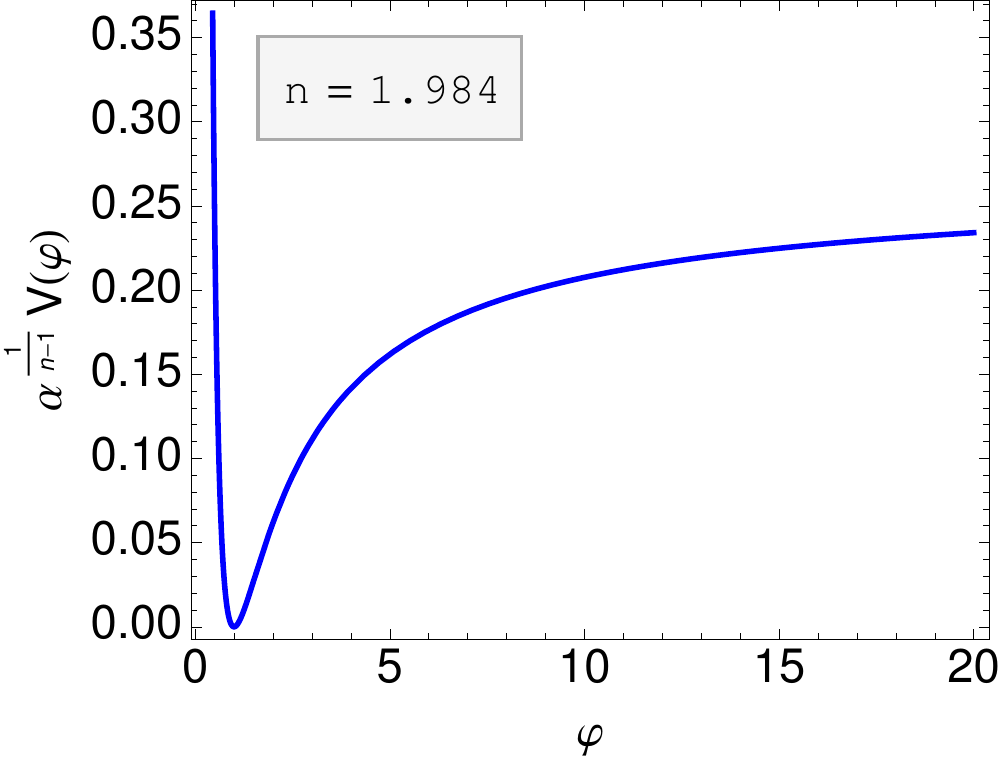}
\hspace{0.1cm}
\includegraphics[height=3.7cm,bb=0 0 288 212]{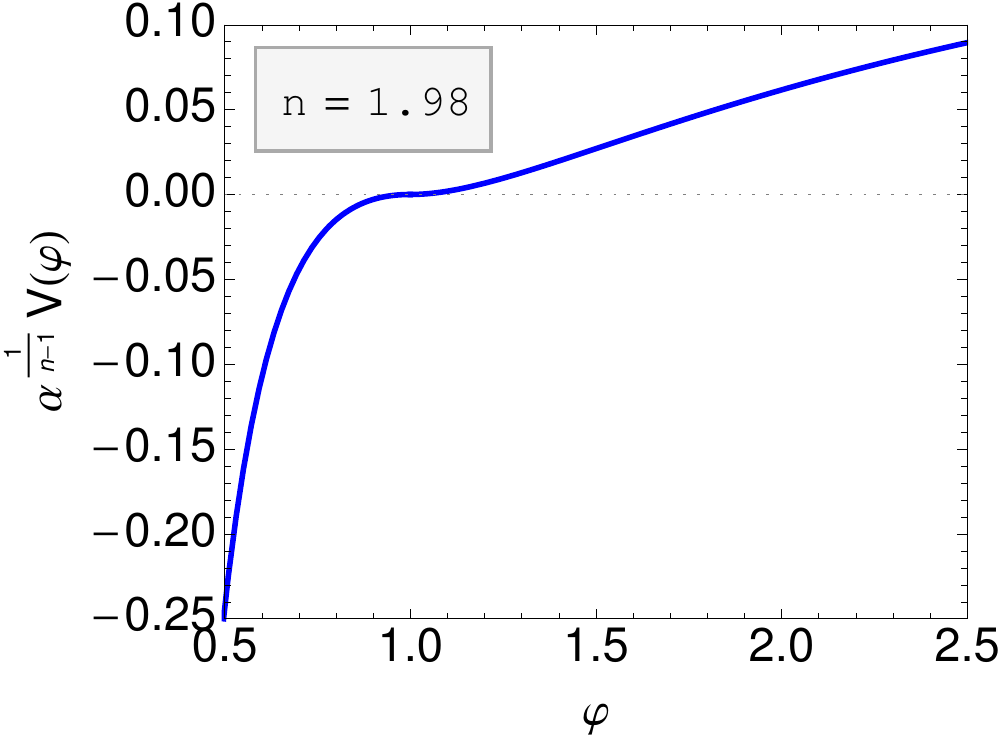}
\hspace{0.1cm}
\includegraphics[height=3.7cm,bb=0 0 288 220]{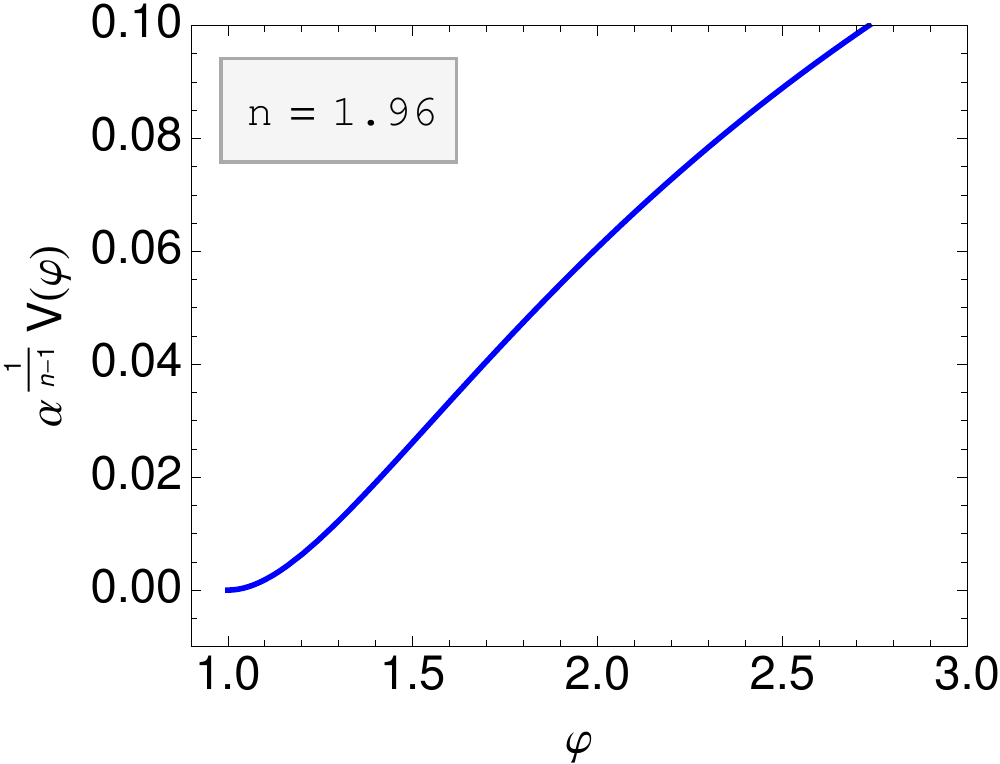}
\caption{\it  All panels show the Einstein frame scalar potential $V(\varphi)$ in the $R + \alpha R^{n}$ model. The left panel describes the case with the stable minimum.  Middle and right panels show the Einstein frame scalar potential $V(\varphi)$ in the case of lack of the minimum. At the right panel the potential obtains complex values for $\varphi<0$.}
\label{fig:Potential_n}
\end{figure}

The Einstein frame scalar potential for $f(R) = R + \alpha R^n$ as a function of $\varphi$ has a following form
\begin{equation}
V(\varphi) = \alpha(n-1)(\alpha n)^{\frac{-n}{n-1}}\left(1-\frac{1}{\varphi}\right)^2(\varphi-1)^{\frac{2-n}{n-1}}\, ,\label{eq:VRn}
\end{equation}
where the last term parametrizes the deviation from the Starobinsky potential. Let us discuss case by case specific choices for the power in the last term in the context of the features of the potential.
\begin{description}
\item{a)} To obtain the real values of the potential for all $\varphi$ together with the stable minimum at $\varphi=1$ one needs
\begin{equation}
\frac{2-n}{n-1} = \frac{2l}{2k+1} \, , \label{eq:condition_minimum}
\end{equation}
where $k,l\in \mathbb{N}$ and $l \leq k$. To satisfy these conditions one need to assume that
\begin{equation}
n = 2(1-2^a 10^{-b})\, ,
\end{equation}
where $a,b\in\mathbb{N}$, $a\geq b$ and $a<b\log_2(10)+\log_2(3-\sqrt{3})-2$. The last condition is needed in order to obtain accelerated expansion of the space-time. This case is presented at the left panel of the Fig. \ref{fig:Potential_n}. For this class of potentials the reheating of the Universe takes place during the oscillations period, after the scalar field reach its minimum.

\item{b)} The real values of the potential for all $\varphi$ and a potential without any minimum is obtained for
\begin{equation}
\frac{2-n}{n-1} = \frac{2l+1}{2k+1} \, ,
\end{equation}
where $k,l\in \mathbb{N}$ and $l < k$. Potential obtains negative values for $\varphi<1$ and it heads to $-\infty$ for $\varphi\to 0$. This case is presented at the middle panel of the Fig. \ref{fig:Potential_n}. Such a potential is non-physical and without additional terms it cannot be used to generate inflation.

\item{c)} In all other cases the potential $V$ becomes imaginary for any $\varphi<1$ and it has no minimum. This case is presented at the right panel of the Fig. \ref{fig:Potential_n}. The inflation ends with the reheating of the Universe, which takes place when the scalar field rolls towards $\varphi=1$. Since there is no oscillation phase such a potential requires strong couplings between the auxiliary field and matter fields.
\end{description}

\subsection{The generation of primordial inhomogeneities}

\begin{figure}[h]
\centering
\includegraphics[height=4.7cm,bb=0 0 288 183]{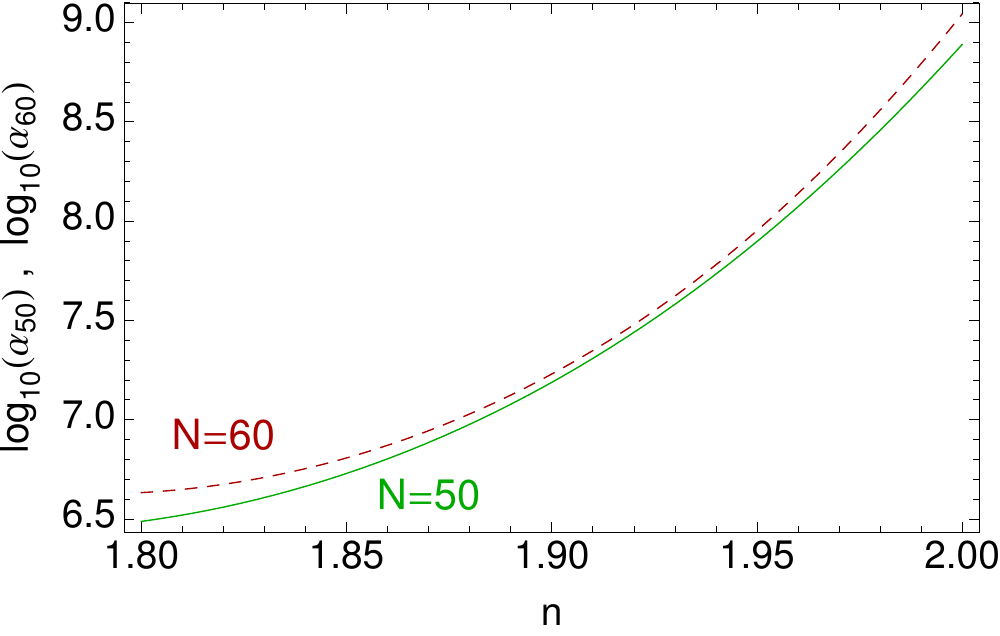}
\hspace{0.5cm}
\includegraphics[height=4.7cm,bb=0 0 288 188]{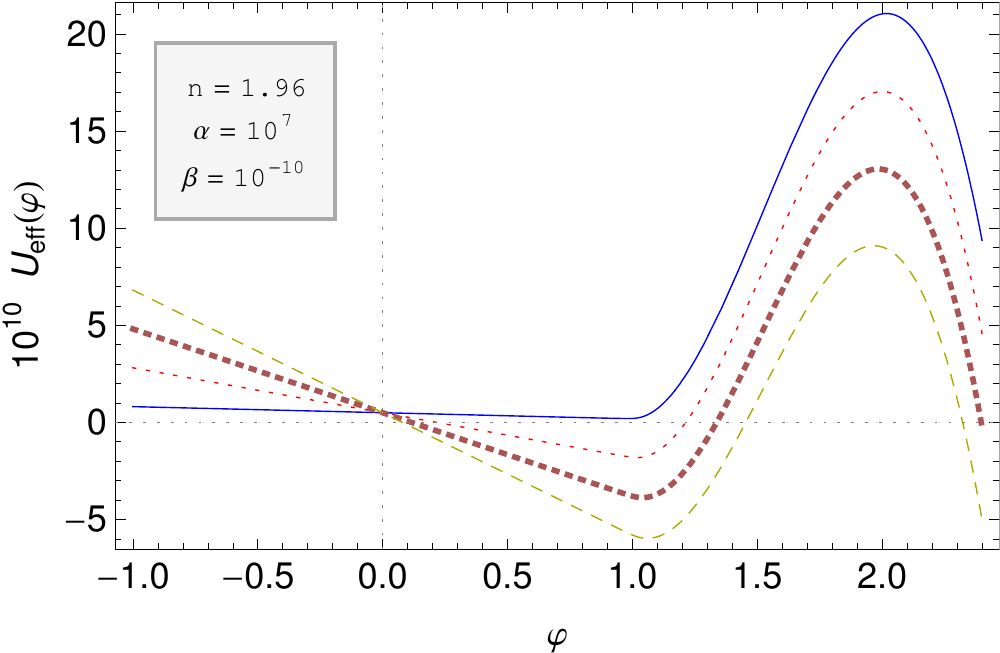}
\caption{\it Left panel: The analytical solution for $\alpha$ as a function of n at the moment of $N=50$ or $N=60$ (solid green and dashed red lines respectively). Right panel: The $U_{eff}$ in the vacuum case (solid blue line) and for the dust with $\rho_M = 2\times10^{10}$ (dotted red line), $\rho_M = 4\times10^{10}$ (thick dotted brown line), $\rho_M = 6\times10^{10}$ (dashed yellow line). The result is presented up to an additive constant. The minimum is getting closer to $\varphi = 1$ as the $\rho_M$ starts to dominate over $U(\varphi)$.}
\label{fig:alpha}
\end{figure}

\begin{figure}[h]
\centering
\includegraphics[height=3.5cm,bb=0 0 288 211]{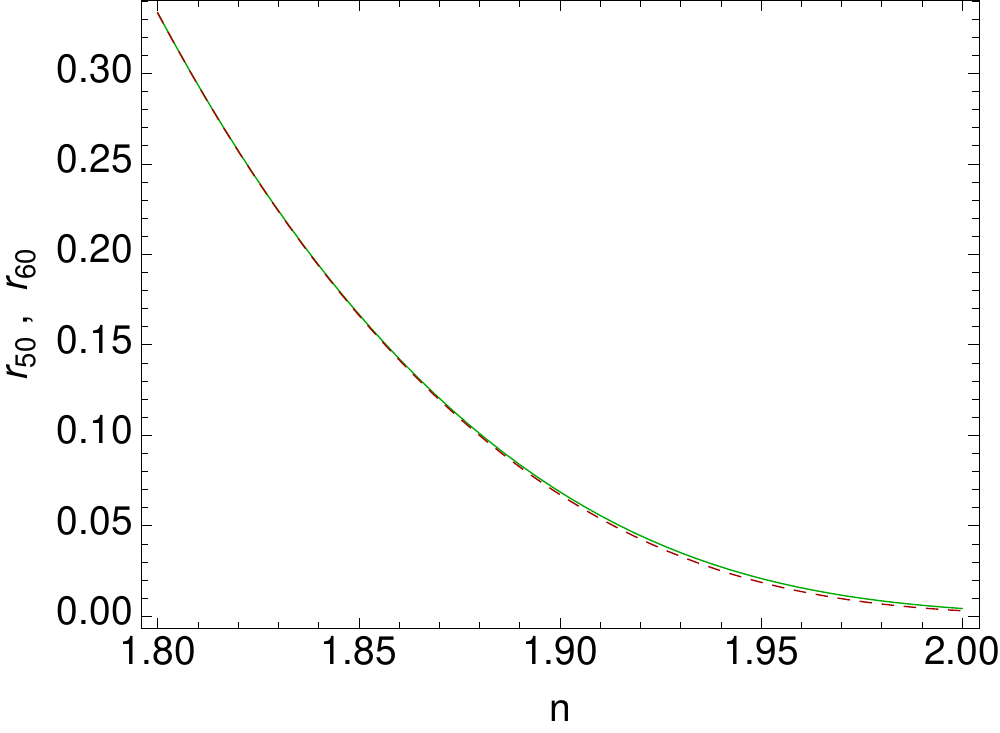}
\hspace{0.3cm}
\includegraphics[height=3.5cm,bb=0 0 288 206]{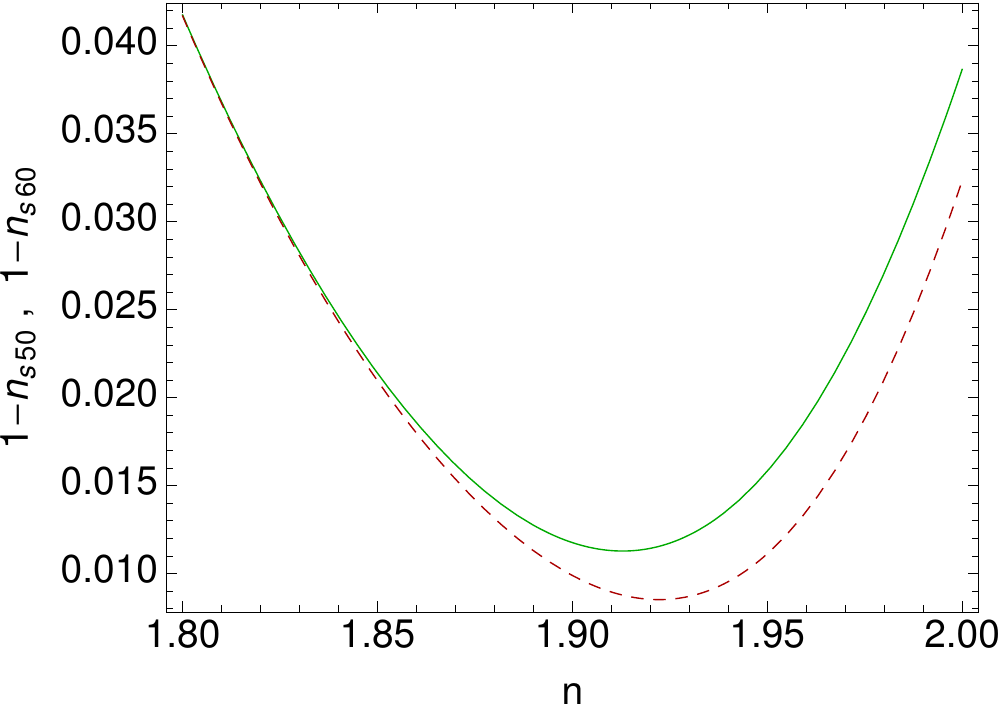}
\hspace{0.3cm}
\includegraphics[height=3.5cm,bb=0 0 291 219]{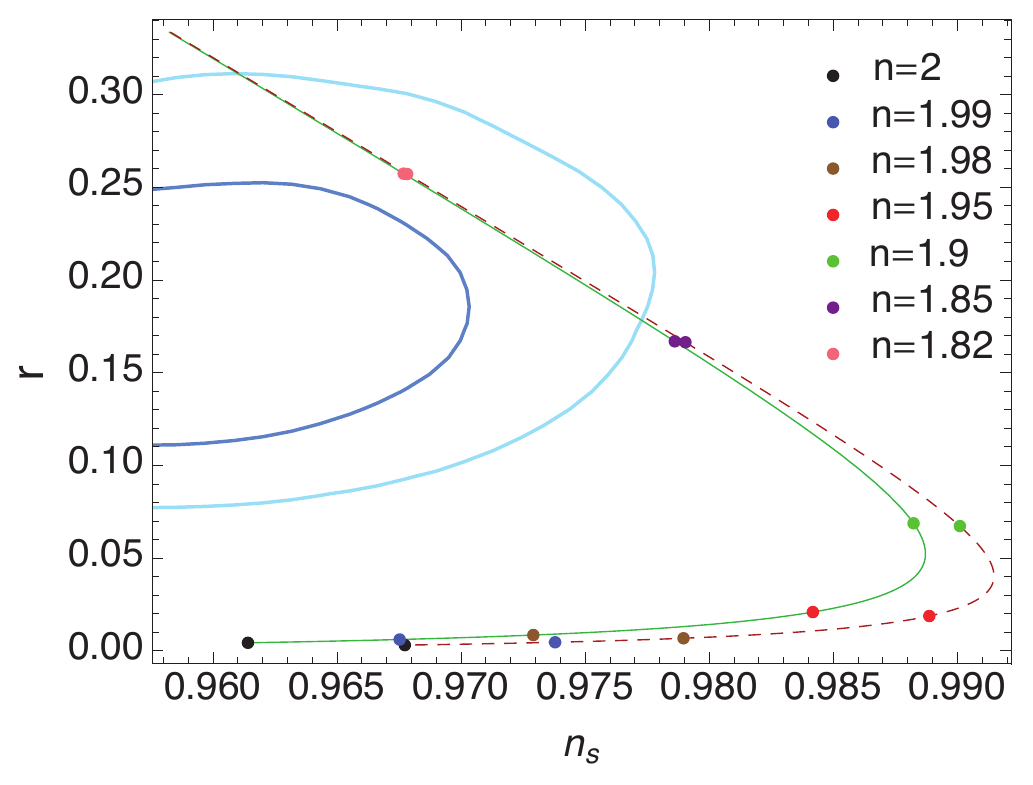}
\caption{\it Left  and middle panels: The analytical solution for $r$ and $n_s$ as a function of n at the moment of $N=50$ or $N=60$ (solid green and dashed red lines respectively). Right panel: The $(r,n_s)$ plane for different $n$. Dark and light blue lines represent $1\sigma$ and $2\sigma$ regions of the BICEP2 results \cite{Ade:2014gua}. The $R + \alpha R^n$ model is inside the $2\sigma$ region for $n\in(1.805,1.845)$. For $n\lesssim 2$ the results are consistent with the PLANCK data.}
\label{fig:perturbations}
\end{figure}

The power spectrum of the Jordan frame primordial curvature perturbations at the super-horizon scales has the following form \cite{DeFelice:2010aj}
\begin{equation}
\mathcal{P}_{\mathcal{R}} \simeq  \frac{2H^2F}{3 \dot{F}^2}\left(\frac{H}{2\pi }\right)^2 \, .\label{eq:PowerSpectrum}
\end{equation}
Let us note that the Eq. (\ref{eq:Fried2}) can be expressed as $\epsilon_H = -\epsilon_F(1-\eta_F)$, so for the slow-roll approximation one finds $\epsilon_H\simeq-\epsilon_F$. This approximation is valid for $2-n\ll1$, since $\epsilon_F=(n-1)\epsilon_H$ for $\epsilon F>1$. Thus, the spectral index $n_s$ and tensor to scalar ratio $r$ are of the form of \cite{DeFelice:2010aj}
\begin{equation}
n_s \simeq 1 -4\epsilon_H + 2{\epsilon_F} - 2\eta_F  \simeq 1 -6\epsilon_H - 2\eta_F \, ,\qquad r\simeq 48 \epsilon_F^2 \simeq {48}\epsilon_H^2 \, .
\end{equation}
The normalisation of primordial inhomogeneities requires that $\mathcal{P}_\mathcal{R}^{1/2}\sim 5\times 10^{-5}$ at the moment of $50$ to $60$ e-folds before the end of inflation. One can use the normalisation of the power spectrum to obtain the realistic values of $\alpha$. From Eq. (\ref{eq:varphiEOM},\ref{eq:FriedBD},\ref{eq:fR},\ref{eq:PowerSpectrum}) in the slow-roll regime one finds the $\alpha = \alpha(n)$, which (for realistic values of $n$) is plotted at the Fig. \ref{fig:alpha}   The $(n_s,r)$ plane (which describes the shape of primordial curvature perturbations) for the $R + \alpha R^{n}$ model have been presented at the Fig. \ref{fig:perturbations}.

For completeness of the description of the $R + \alpha R^{n}$ model we present in the Appendix \ref{app:osc} the analysis of the final stage of the evolution of the field near the minimum, in cases when the minimum does exist.

\section{The $R + \alpha R^{n} - \beta R^{2-n}$ model} \label{sec:R^1+m}

Deficiencies of the model discussed in the previous Section can be bypassed by considering further modification of the gravitational Lagrangian. To generate a minimum for the scalar potential and (as we will show) to generate dark energy as a relict of inflation let us consider more complicated form of $f(R)$, namely
\begin{equation}
f(R) = R + \alpha R^{n} - \beta R^{2-n} \, ,
\end{equation}
where $\alpha$ and $\beta$ are positive constants and $n$ satisfies the Eq. (\ref{eq:nin}). Let us require $\alpha\gg1$,  $\beta\ll1$ and $\alpha\beta\ll1$. This means that the $\alpha R^{n}\gg R\gg\beta R^{2-n}$ during inflation, so the results obtained in the sec. \ref{sec:R^n} are still valid. From 
\begin{equation}
 \varphi = F = 1+\alpha n R^{n-1} - \beta(2-n) R^{1-n}
\end{equation}
one finds
\begin{equation}
R(\varphi) = \left(\frac{\sqrt{4 (2-n)n \alpha  \beta +(\varphi -1)^2}+\varphi -1}{2n \alpha }\right)^{\frac{1}{n-1}}
\end{equation}
Let us note that {$R > 0$} for any value of $\varphi$. {For $\varphi - 1 \gg \alpha \beta$ one obtains
\begin{equation}
R(\varphi) \simeq \left(\frac{\varphi-1}{n\alpha}\right)^{\frac{1}{n-1}} \quad \Rightarrow \quad \frac{\alpha R^n}{\beta R^{2-n}}\simeq \frac{1}{\alpha\beta}\left(\frac{\varphi-1}{n}\right)^2 \gg 1 \, .\label{eq:infltertmm+1}
\end{equation}
Thus, the $\beta R^{2-n}$ term does not have any influence on inflation and generation of the large scale structure of the universe.} 
\\*

\begin{figure}[h]
\centering
\includegraphics[height=4.75cm,bb=0 0 288 189]{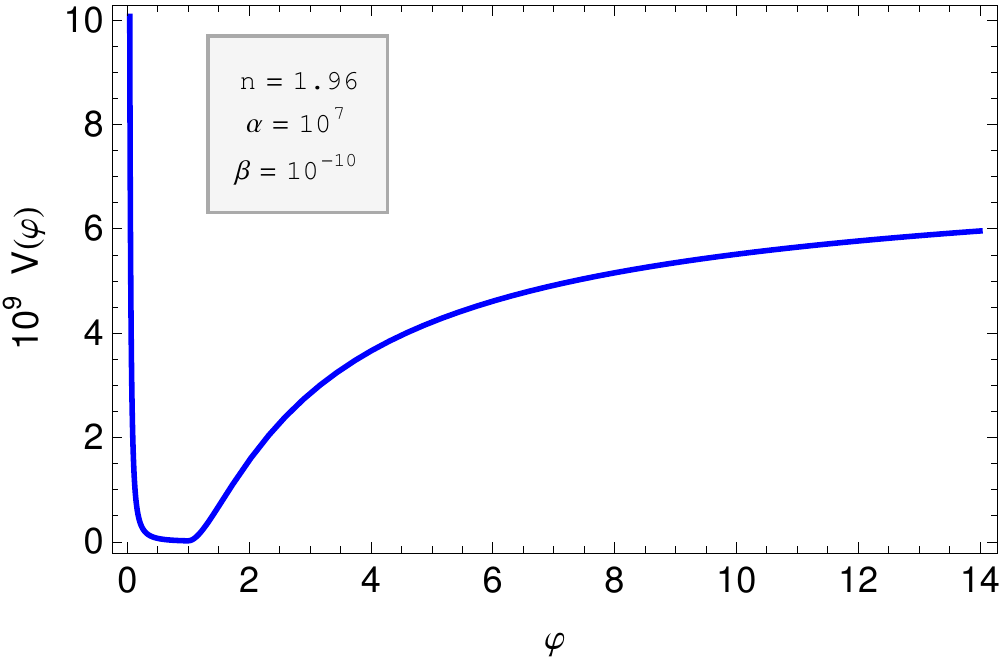}
\hspace{0.5cm}
\includegraphics[height=4.75cm,bb=0 0 288 185]{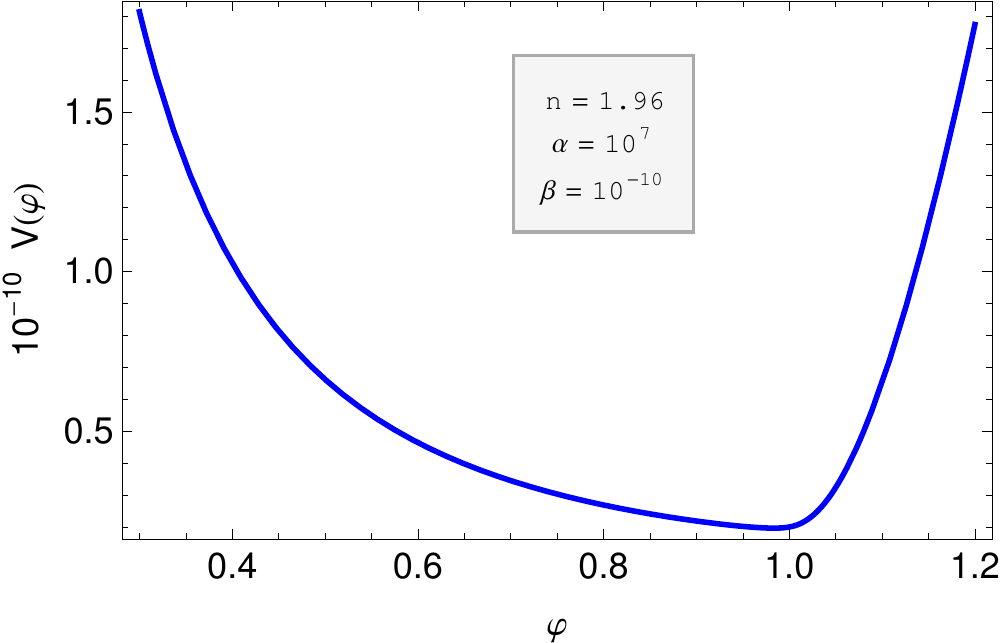}
\caption{\it  Both panels show the Einstein frame scalar potential $V(\varphi)$ and its minimum in the $R + \alpha R^{n} - \beta R^{2-n}$ model. Note that $V(\varphi_{min})\sim 2\times 10^{-11}$, which gives the non-zero vacuum energy.}
\label{fig:Potential_m+1}
\end{figure}

The Jordan and Einstein frames scalar potentials {look} as follows
\begin{equation}
U(\varphi) = \frac{1}{2} (n-1)\left(\alpha  R^{ n}(\varphi )+\beta  R^{2-n}(\varphi )\right) \, ,\quad  V = \frac{1}{\varphi^2}U
\end{equation}
The Einstein frame scalar potential has a minimum at
\begin{equation}
R_{min}=\left(\frac{\sqrt{1+4 (2-n) n \alpha  \beta }-1}{2(2-n) \alpha }\right)^{\frac{1}{n-1}}\simeq (n \beta )^{\frac{1}{n-1}} \left(1-\frac{(2-n) n \alpha  \beta }{n-1}\right) \, ,
\end{equation}
where the last term is the Taylor expansion with respect to $beta$. The minimum is slightly shifted with respect to  $R=0$, which is the GR vacuum case. Hence, this model predicts some amount of vacuum energy. Around the minimum one finds
\begin{equation}
\alpha R_{min}^{n-1} \sim n \alpha \beta \ll 1 \, ,\qquad \beta R_{min}^{1-n} \sim \frac{1}{n} \sim O(1) \, ,
\end{equation}
which means that the inflationary term is negligible and $R\sim \beta R^{2-n}$. The existence of the $\alpha R^{n}$ term is still important, since it provides the minimum and real values of a scalar potential for all $\varphi$. Let us clarify that $R_{min}$ is not the minimal value of $R$. The Ricci scalar has no minimum, its minimal value is equal to 0 (at the $\varphi\to-\infty$ limit) and it continuously grows with $\varphi$. The value of $V$ at the minimum for small values of $\beta$ reads
\begin{equation}
V(\varphi_{min})\simeq \frac{n}{8 (n-1)^2} (n \beta )^{\frac{1}{n-1}} \left(n-1-n^2 \alpha  \beta \right) \sim \frac{1}{2}\beta^{\frac{1}{n-1}} \, ,\label{eq:Vmin}
\end{equation}
where $\varphi_{min}:=F(R_{min})\simeq \frac{2}{n} (n-1) (1+2 n \alpha  \beta )$ is the value of $\varphi$ at the minimum of $V$. The Eq. \ref{eq:Vmin} means that the energy density of the DE shall be of order of $\beta^{\frac{1}{n-1}}$ and that we need $\beta\lll1$ to fit the dark energy data. An example of a potential and its minimum are plotted at the Fig. \ref{fig:Potential_m+1}. The existence of  a stable minimum is one of the main differences between this model and a $R-\beta R^{2-n}$ dark energy model, in which the auxiliary field rolls down towards negative $\varphi$ for small energies \cite{Amendola:2006we}. The minimum of the Einstein frame potential (visible at both panels of the Fig. \ref{fig:Potential_m+1}) prevents the $\varphi$ from obtaining negative values for any solution with inflationary initial conditions.

\subsection{Conditions for the $f(R)$ dark energy}

Let us check whether this model satisfies conditions pointed out in the Ref. \cite{DeFelice:2010aj}. The $R_0$ denoted the value of the Ricci scalar today

\begin{description}
\item 1) To avoid the ghost state one needs
\begin{equation}
F>0 \qquad \text{for}\qquad R\geq R_0 \, ,
\end{equation}
which means that 
{\begin{equation}
R_0>R(\varphi=0)=\left(\frac{\sqrt{1+4n \alpha  \beta (2-n)}-1}{2\alpha  n}\right)^{\frac{1}{n-1}} \, .
\end{equation}
This condition is satisfied for any $\varphi$ with initial condition $\varphi>0$, due to the existence of the minimum of the Einstein frame scalar potential. This is an advantage of this model comparing to the $f(R) = R - \beta R^{2-n}$ one, in which $F<0$ in late times.}
\item 2) To avoid the negative mass square for a scalar field degree of freedom one needs
\begin{equation}
F'>0 \qquad \text{for}\qquad R\geq R_0\, ,
\end{equation}
which means that $R_0>0$. This condition is obviously satisfied, since the universe at the present times is filled with DE and dust.
\item 3) To satisfy consistency with local gravity constraints one needs
\begin{equation}
f(R) \to R - 2\Lambda \qquad \text{for}\qquad R\geq R_0 \, ,
\end{equation}
After the $\varphi$ field is stabilised in its minimum it produces the vacuum energy, which is a source of $\Lambda$. The non-zero values of $R$ comes from radiation and dust produced during the reheating of the universe, as well as from the non-zero minimal value of the Ricci scalar.
\item 4) For the stability and the presence of a late-time de Sitter solution one needs
\begin{equation}
0<\frac{RF'}{F}<1 \qquad \text{at}\qquad r:=-\frac{RF}{f}=-2
\end{equation}
The $r=-2$ happens for
{\begin{equation}
R=R_r=\left(\frac{\sqrt{1+4n \left(2-n\right) \alpha  \beta }-1}{2(2-n) \alpha }\right)^{\frac{1}{n-1}} \simeq (n \beta )^{\frac{1}{n-1}} \, .
\end{equation}}
At $R=R_r$ the $RF'/F$ takes the form of
\begin{equation}
\frac{RF'}{F}(r=-2) = \frac{1}{2+8 \alpha  \beta }\left(1+4 \left(n^2-2n+2\right) \alpha  \beta -(n-1) \sqrt{1+4 n\left(2-n\right) \alpha  \beta }\right) \, ,
\end{equation}
which lies in the range of $(0,1)$ for considered regimes of $\alpha$ and $\beta$. To see that clearly let us perform a Tylor expansion of the $RF'/F$ with respect to $\alpha\beta$. Then one obtains
\begin{equation}
\frac{RF'}{F}(r=-2)\simeq \frac{1}{2}(2-n)+(n-1) n^2 \alpha \beta
\end{equation}
Thus, this model satisfies conditions for a viable DE model.

\end{description}

\section{Numerical analysis of the dark energy model} \label{sec:DE}

\begin{figure}[h]
\centering
\includegraphics[height=4.45cm,bb=0 0 288 173]{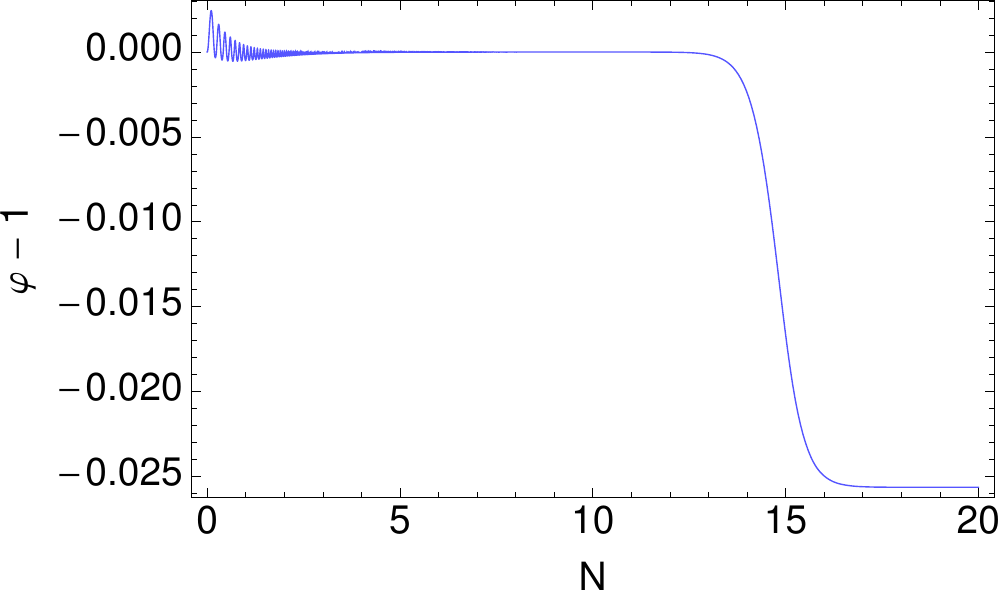}
\hspace{0.5cm}
\includegraphics[height=4.45cm,bb=0 0 288 177]{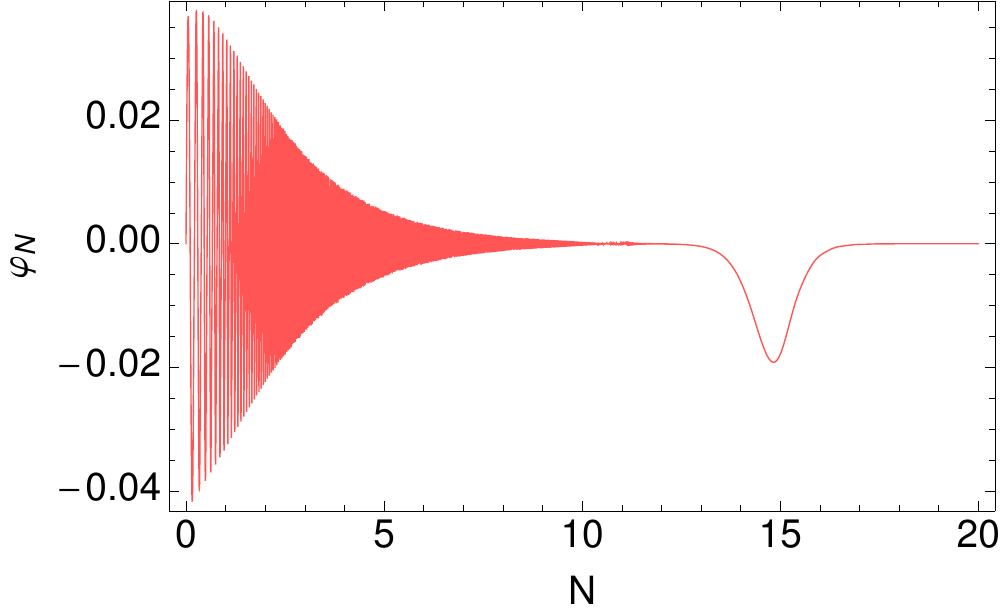}
\caption{\it Numerical results of the evolution of the Brans-Dicke field $\varphi$ and its derivative with respect to $N$ as a function of $N$. We have assumed $n=1.95$, $\alpha = 2\times 10^8$, $\beta = 10^{-30}$, $\varphi(0) = 1$ and $\varphi_N(0) = 0$. The $\rho_M$ is the dust with $\rho_I\sim 10^{-12}$. The choice of initial conditions and parameters of the model is rather unrealistic (to big $\beta$ and $\rho_I$), but it illustrates the way of the field to its minimum.}
\label{fig:phiDE}
\end{figure}

\begin{figure}[h]
\centering
\includegraphics[height=4.5cm,bb=0 0 288 182]{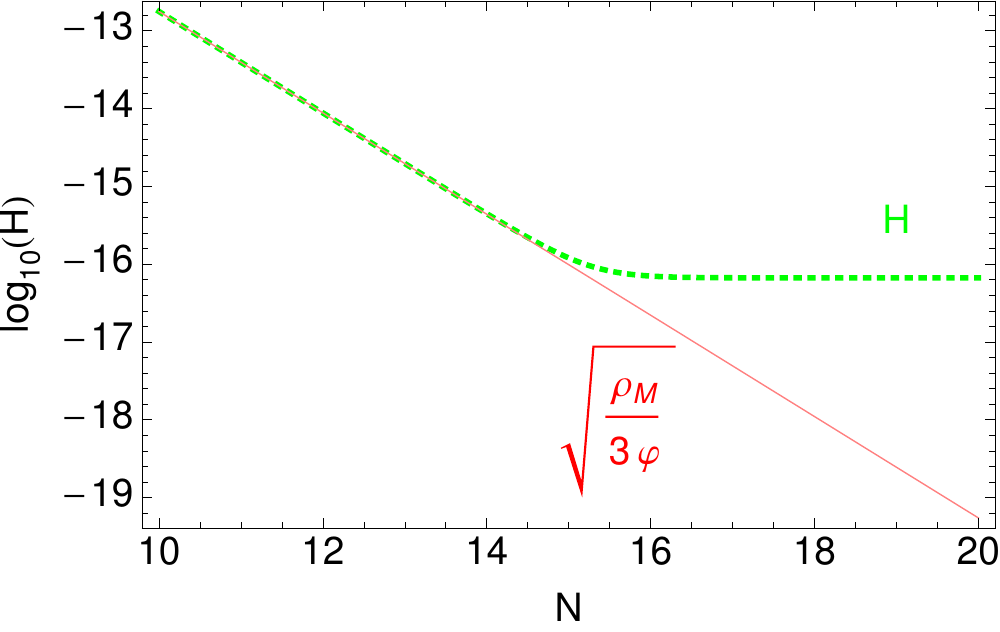}
\hspace{0.5cm}
\includegraphics[height=4.5cm,bb=0 0 288 180]{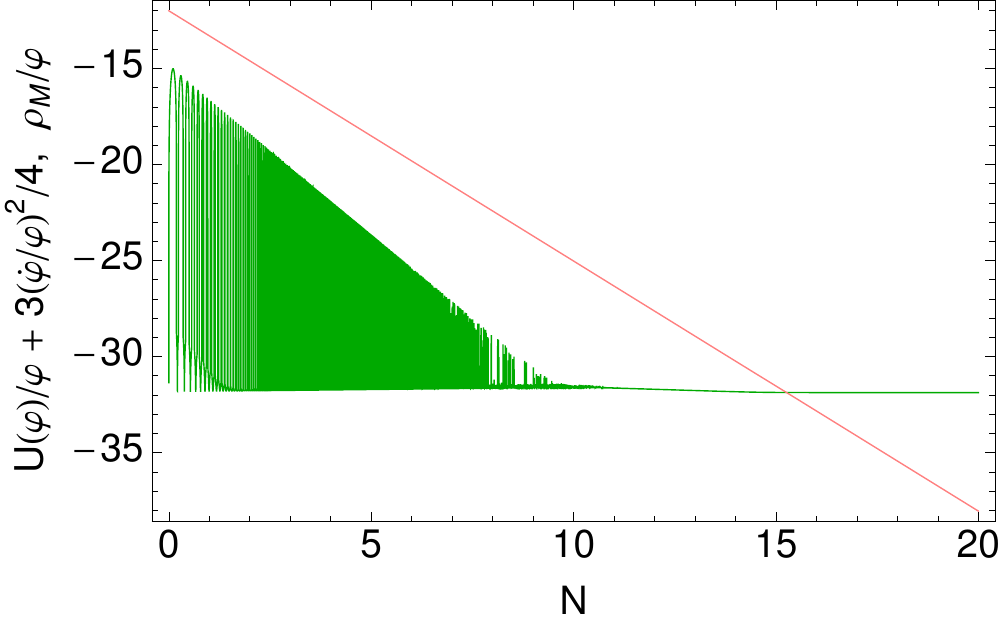}
\caption{\it Left panel: numerical result of the evolution of the Hubble parameter (green dashed line) and $\sqrt{\rho_M/3\varphi}$ (red solid line) as a function of time. Note that the vacuum energy of the Brans-Dicke field starts to dominate when the field reaches it's minimum. Right panel: numerical result for the evolution of dust and auxiliary field's energy density (red and green lines respectively). The energy density of the $\varphi$ obtains the barotropic parameter $\omega_\varphi\simeq-1$ about 10 e-folds before the DE domination period.}
\label{fig:HDE}
\end{figure}

The non-zero value of the Einstein frame potential at the minimum rises a possibility of obtaining a realistic solution to the dark energy problem. To analyse low energy solutions of the Jordan frame equations of motion let us use the number of e-folds (defined by $N:=\log(a)$) as a time variable. Then Eq. (\ref{eq:varphiEOM},\ref{eq:FriedBD}) read
\begin{eqnarray}
H^2(\varphi_{NN}+3\varphi_N) + H_NH\varphi_N+\frac{2}{3}(\varphi U_\varphi-2U) &=& \frac{1}{3}(\rho_M-3p_M)\, ,\\
H^2 &=& \frac{\rho_M + U}{3\left(\varphi + \varphi_N\right)} \, ,
\end{eqnarray}
where the index ``$ _N$'' denotes the derivative with respect to $N$. Since in the Jordan frame the Eq. (\ref{eq:cont}) is satisfied one finds $\rho_M=\rho_I e^{-3(1+w)N}$, where $w=p_M/\rho_M$ is a barotropic parameter. After the inflation $\varphi$ oscillates around $\varphi_{\min}$ and reheats the universe by the particle production. Thus, after oscillations one obtains the radiation domination era, for which $w=1/3$ and $\rho_M - 3p_M = 0$. The radiation increases the cosmic friction term but does not contributes to the $U_{eff}$, so the field is not shifted from the minimum. However, during the dust domination era the $U_{eff}$ is modified and $\varphi$ oscillates around $\varphi=1$. The evolution of $\varphi$ and $\varphi_N$ during the dust/DE domination era is presented at the Fig. \ref{fig:phiDE}. The evolution of the Hubble parameter and $\sqrt{\rho_M/3}$ is plotted at the Fig. \ref{fig:HDE}. We have assumed that the field starts from the $\varphi\simeq 1$ (which is the GR limit of the theory), but numerical analysis shows that the late-time results do not depend on initial conditions. For instance the initial value of the field in the Einstein frame minimum the only difference is slightly longer period of oscillations around $\varphi=1$.  As long as the $\rho_M$ dominates over the energy density of the Brans-Dicke field the field oscillates and stabilises above the $\varphi_{min}$. During that period $|\varphi-1|\ll1$ and the $R$ term in $f(R)$ dominates. Thus one recovers the GR limit of the theory. When the dust becomes subdominant the $\varphi$ rolls to its minimum and one obtains the Dark Energy with the barotropic parameter $\omega=-1$.
\\*

As shown in the Fig. \ref{fig:HDE} the Hubble parameter obtains the constant value when the auxiliary field is in its minimum. The $H = const$ implies that $R=12H^2 = 4\rho_\Lambda$, where $\rho_\Lambda$ is the DE energy density. While $\varphi$ is in its minimum one finds $R = R_{min}$, so at the late times one finds
\begin{equation}
\rho_\Lambda = R_{min}/4 \simeq \frac{1}{4}(n\beta)^{\frac{1}{n-1}} \qquad \Rightarrow \qquad \beta \simeq \frac{1}{n}\left(4\rho_\Lambda\right)^{n-1}\, .\label{eq:DEofbeta}
\end{equation}
The $\rho_\Lambda$ becomes constant even before the DE domination era. Thus, the Eq. \ref{eq:DEofbeta} shall be satisfied at the present time.

\section{Conclusions} \label{sec:concl}

In this note we have demonstrated how to extend the $f(R) = R + \alpha R^n$ model to combine successfully both features of the observable Universe in a single framework: we show how to obtain successful inflation and the non-zero residual value of the Ricci scalar in an extension of the Starobinsky model. Our results can be easily consistent with PLANCK or BICEP2 data for appropriate choices of the value of $n$. In this case the Einstein frame potential of the auxiliary field has a minimum only when $(2-n)/(n-1)$ is of the form of a fraction with even and odd numbers in the numerator and the denominator respectively. The potential has a minimum at $\varphi=1$ and $V(\varphi=1)=0$, which means that there is no vacuum energy. For all other forms of $(2-n)/(n-1)$ the minimum does not exist and the potential goes to $-\infty$ or becomes complex for $\varphi<1$.
\\*

In the section \ref{sec:R^1+m} we have generalised this model into $R + \alpha R^n - \beta R^{2-n}$, with $\alpha\gg1$ and $\beta\ll1$. In this case the Ricci scalar is always bigger than zero. The Einstein frame scalar potential is real for all $\varphi$ and it has a minimum for all $n$. The potential has non-zero value at the minimum, which may become a source of DE. The value of the
 parameter $\alpha$ is set by the normalisation of primordial inhomogeneities, while the value of the parameter $\beta$ can be read from the measured value of the present DE energy density.
\\*

In the section \ref{sec:DE} we have performed numerical analysis of the late-time evolution of the $R + \alpha R^n - \beta R^{2-n}$ model with dust employed as a matter field. During the radiation domination era the $\rho_M-3p_M = 0$, so the effective potential in the Jordan frame obtains its vacuum form. Thus the field holds $\varphi=\varphi_{min}$. During the dust domination era one finds $\varphi = F\simeq 1$, which corresponds to the GR limit of the theory. When matter starts to be subdominant the $\varphi$ rolls to its minimum in $\varphi=\varphi_{min}\lesssim1$. Even before that moment the energy density of the auxiliary field becomes constant and the energy density of the $\varphi$ evolves like DE with barotropic parameter $\omega_\varphi=-1$. From the present value of the DE energy density we have found the value of the  parameter $\beta$ to be $(4\rho_\Lambda)^{n-1}/n$.

\section{Acknowledgements}

We would like to thank prof. Ma for useful discussions.\\ 
This work has been supported by National Science Centre under research grant\\ DEC-2012/04/A/ST2/00099 and partially by research grant DEC-2011/01/M/ST2/02466,  by NSFC (No.11235003) and the Fundamental Research Funds for the Central University of China under Grant No.2013ZM107. M.A. would also like to acknowledge China Postdoctoral Science Foundation for financial support.

\appendix

\section{Oscillations around the minimum} \label{app:osc}

Let us assume that the condition (\ref{eq:condition_minimum}) is satisfied. Then the inflation ends with the period of oscillations of the scalar field around the minimum of its potential. To obtain the approximate form of the potential $V$ from the Eq. (\ref{eq:VRn}) around the minimum $\varphi=1$ one shall express $V$ in term of the Einstein frame field $\phi = \sqrt{3/2}\log(\varphi)$ and perform a Tylor expansion around $\phi=0$. Then the leading term is equal to
\begin{equation}
V(\phi)\simeq 6^{\frac{-n}{2(n-1)}} 4(n-1) \alpha  (n \alpha )^{\frac{n}{n-1}} \phi ^{\frac{n}{n-1}}\, .\label{eq:VRn_app}
\end{equation}
During the oscillation phase one obtains 
\begin{equation}
<\phi_{\tilde{t}}^2> \simeq <V_\phi \phi> \, ,\label{eq:avarage}
\end{equation}
where { $< \, >$ }denotes the average value over one period of oscillations. Thus, the effective barotropic parameter in the Einstein frame (denotes as $\omega_\phi$) is equal to
\begin{equation}
\omega_\phi=\frac{<p>}{<\rho>} = \frac{\frac{1}{2}<V_\phi \phi>-<V>}{\frac{1}{2}<V_\phi \phi>+<V>} \, ,
\end{equation}
where $V_\phi = \frac{dV}{d\phi}$. For the potential of the form of (\ref{eq:VRn_app}) one finds 
\begin{equation}
\omega_\phi = \frac{2-n}{3n-2} \, .
\end{equation}
Thus, during the oscillation phase the Einstein frame scale factor is proportional to
\begin{equation}
\tilde{a} = \tilde{t}^{\frac{2}{3(1+\omega_\phi)}} = \tilde{t}^{\frac{3n-2}{3 n}}
\end{equation}
From the first Firedmann equation in the Einstein frame one obtains
\begin{equation}
<3\tilde{H}^2> =  <\frac{1}{2}\phi_{\tilde{t}}^2> + <V> = \frac{1}{2}<V_\phi \phi>+<V>\, ,\label{eq:FriedAv}
\end{equation}
where $\tilde{H}:=\tilde{a}_{\tilde{t}}/\tilde{a}$ is the Einstein frame Hubble parameter. From Eq. (\ref{eq:avarage},\ref{eq:FriedAv}) one finds
\begin{equation}
<\phi> \simeq 6^{\frac{1}{n}-\frac{1}{2}}(\text{n$\alpha $})^{-\frac{1}{n}}\left(\frac{3 n-2 }{t^2n^3 \alpha ^2}\right)^{\frac{n-1}{n}}, \qquad <\varphi> = \exp\left[\frac{1}{3}\left(\frac{6}{\text{n$\alpha $}}\right)^{\frac{1}{n}}\left(\frac{3 n-2 }{t^2n^3 \alpha ^2}\right)^{\frac{n-1}{n}}\right]\, .
\end{equation}
By definition the Jordan frame scale factor is equal to $a = \tilde{a}\varphi^{-1/2}$, which gives us the evolution of $a$ as a function of the Einstein frame time variable.


\begin{thebibliography}{99}


\bibitem{LR99}D. H. Lyth and A. Riotto, 
Phys. Rept. 314, 1 (1999).

\bibitem{Liddle}A. R. Liddle, 
New Astronomy Reviews, 45, 235-253 (2001).

\bibitem{MR11}A. Mazumdar and J. Rocher, 
Phys. Rept. 497, 85 (2011).

\bibitem{Starobinsky:1980te}
  A.~A.~Starobinsky,
  Phys.\ Lett.\ B {\bf 91} (1980) 99.

\bibitem{Ade:2014xna}
  P.~A.~R.~Ade {\it et al.}  [BICEP2 Collaboration],
  arXiv:1403.3985 [astro-ph.CO].
  
\bibitem{Ade:2014gua}
  P.~A.~RAde {\it et al.}  [BICEP2 Collaboration],
  arXiv:1403.4302 [astro-ph.CO].

\bibitem{Codello:2014sua}
  A.~Codello, J.~Joergensen, F.~Sannino and O.~Svendsen,
  arXiv:1404.3558 [hep-ph].


\bibitem{DeFelice:2010aj}
  A.~De Felice and S.~Tsujikawa,
  Living Rev.\ Rel.\  {\bf 13} (2010) 3
  [arXiv:1002.4928 [gr-qc]].


\bibitem{Ben-Dayan:2014isa}
  I.~Ben-Dayan, S.~Jing, M.~Torabian, A.~Westphal and L.~Zarate,
  arXiv:1404.7349 [hep-th].


\bibitem{Sebastiani:2013eqa}
  L.~Sebastiani, G.~Cognola, R.~Myrzakulov, S.~D.~Odintsov and S.~Zerbini,
  Phys.\ Rev.\ D {\bf 89} (2014) 023518
  [arXiv:1311.0744 [gr-qc]].
  
\bibitem{Perlmutter:1999jt}
  S.~Perlmutter, M.~S.~Turner and M.~J.~White,
  Phys.\ Rev.\ Lett.\  {\bf 83} (1999) 670
  [astro-ph/9901052].
  
  
\bibitem{Spergel:2006hy}
  D.~N.~Spergel {\it et al.}  [WMAP Collaboration],
  Astrophys.\ J.\ Suppl.\  {\bf 170} (2007) 377
  [astro-ph/0603449].
  
  
\bibitem{Ade:2013zuv}
  P.~A.~R.~Ade {\it et al.}  [Planck Collaboration],
  arXiv:1303.5076 [astro-ph.CO].



\bibitem{Capozziello:2002rd}
  S.~Capozziello,
  Int.\ J.\ Mod.\ Phys.\ D {\bf 11} (2002) 483
  [gr-qc/0201033].
  
  
\bibitem{Copeland:2006wr}
  E.~J.~Copeland, M.~Sami and S.~Tsujikawa,
  Int.\ J.\ Mod.\ Phys.\ D {\bf 15} (2006) 1753
  [hep-th/0603057].


\bibitem{Nojiri:2003ft}
  S.~'i.~Nojiri and S.~D.~Odintsov,
  Phys.\ Rev.\ D {\bf 68} (2003) 123512
  [hep-th/0307288].
  
  
  
\bibitem{Amendola:2006we}
  L.~Amendola, R.~Gannouji, D.~Polarski and S.~Tsujikawa,
  Phys.\ Rev.\ D {\bf 75} (2007) 083504
  [gr-qc/0612180].






\end{thebibliography}
\end{document}